\documentclass[%
reprint,
superscriptaddress,
 amsmath,amssymb,
 aps,
]{revtex4-2}

\usepackage{graphicx}% Include figure files
\usepackage{dcolumn}% Align table columns on decimal point
\usepackage{bm}% bold math
\usepackage[caption=false]{subfig}
\usepackage{siunitx, soul, xcolor}
\usepackage{adjustbox}
\usepackage{gensymb}
\usepackage{color}

\usepackage{xparse,xcoffins}

\ExplSyntaxOn
\NewCoffin\imagecoffin
\NewCoffin\labelcoffin

\keys_define:nn { miguel/label }
 {
  label   .tl_set:N = \l_miguel_label_tl,
  labelbox .bool_set:N = \l_miguel_label_box_bool,
  labelbox .default:n = true,
  fontsize .tl_set:N = \l_miguel_label_size_tl,
  fontsize .initial:n = \footnotesize,
  pos .choice:,
  pos/nw .code:n = \tl_set:Nn \l_miguel_label_pos_tl { left,up },
  pos/ne .code:n = \tl_set:Nn \l_miguel_label_pos_tl { right,up },
  pos/sw .code:n = \tl_set:Nn \l_miguel_label_pos_tl { left,down },
  pos/se .code:n = \tl_set:Nn \l_miguel_label_pos_tl { right,down },
  pos/n .code:n = \tl_set:Nn \l_miguel_label_pos_tl { hc,up },
  pos/w .code:n = \tl_set:Nn \l_miguel_label_pos_tl { left,vc },
  pos/s .code:n = \tl_set:Nn \l_miguel_label_pos_tl { hc,down },
  pos/e .code:n = \tl_set:Nn \l_miguel_label_pos_tl { right,vc },
  pos .initial:n = nw,
  unknown .code:n   = \clist_put_right:Nx \l_miguel_label_clist
                       { \l_keys_key_tl = \exp_not:n { #1 } }
 }
\clist_new:N \l_miguel_label_clist
\box_new:N \l_miguel_label_box
\box_new:N \l_miguel_label_image_box

\NewDocumentCommand{\xincludegraphics}{O{}m}
 {
  \group_begin:
  \tl_clear:N \l_miguel_label_tl
  \clist_clear:N \l_miguel_label_clist
  \keys_set:nn { miguel/label } { #1 }
  \tl_if_empty:NTF \l_miguel_label_tl
   {
    \miguel_includegraphics:Vn \l_miguel_label_clist { #2 }
   }
   {
    \SetHorizontalCoffin\imagecoffin
     {
      \miguel_includegraphics:Vn \l_miguel_label_clist { #2 }
     }
    \SetHorizontalCoffin\labelcoffin
     {
      \raisebox{\depth}
       {
        \bool_if:NTF \l_miguel_label_box_bool
         { \fcolorbox{white}{white}{\l_miguel_label_size_tl\l_miguel_label_tl} }
         { \l_miguel_label_size_tl\l_miguel_label_tl }
       }
     }
    \SetVerticalPole\imagecoffin{left}{-2pt+\CoffinWidth\labelcoffin/2}
    \SetVerticalPole\imagecoffin{right}{\Width-3pt-\CoffinWidth\labelcoffin/2}
    \SetHorizontalPole\imagecoffin{up}{\Height+5pt-\CoffinHeight\labelcoffin/2}
    \SetHorizontalPole\imagecoffin{down}{3pt+\CoffinHeight\labelcoffin/2}
    \use:x{\JoinCoffins\imagecoffin[\l_miguel_label_pos_tl]\labelcoffin[vc,hc]} 
    \TypesetCoffin\imagecoffin
   }
   \group_end:
 }
\NewDocumentCommand{\setlabel}{m}
 {
  \keys_set:nn { miguel/label } { #1 }
 }

\cs_new_protected:Nn \miguel_includegraphics:nn
 {
  \includegraphics[#1]{#2}
 }
\cs_generate_variant:Nn \miguel_includegraphics:nn { V }

\ExplSyntaxOff

\begin{document}
\preprint{APS/123-QED}

\title{Magneto-Optical Detection of the Orbital Hall Effect in Chromium}

\author{Igor Lyalin}
\affiliation{Department of Physics, The Ohio State University, Columbus, Ohio 43210, United States}
\author{Sanaz Alikhah}
\affiliation{Department of Physics and Astronomy,  Uppsala University, P.\,O.\ Box 516, SE-75120 Uppsala, Sweden}
\author{Marco Berritta}
\affiliation{Department of Physics and Astronomy,  Uppsala University, P.\,O.\ Box 516, SE-75120 Uppsala, Sweden}
\affiliation{Department of Physics and Astronomy, University of Exeter, Stocker Road, Exeter EX4 4QL, United Kingdom}
\author{Peter M. Oppeneer}
\affiliation{Department of Physics and Astronomy,  Uppsala University, P.\,O.\ Box 516, SE-75120 Uppsala, Sweden}
\author{Roland K. Kawakami}
\affiliation{Department of Physics, The Ohio State University, Columbus, Ohio 43210, United States}

\date{\today}

\begin{abstract}
The orbital Hall effect has been theoretically predicted but its direct observation is a challenge.
Here, we report the magneto-optical detection of current-induced orbital accumulation at the surface of a light 3$d$ transition metal, Cr. 
The orbital polarization is in-plane, transverse to the current direction, and scales linearly with current density, consistent with the orbital Hall effect.
Comparing the thickness-dependent magneto-optical measurements with \textit{ab initio} calculations, we estimate an orbital diffusion length in Cr of $6.6\pm 0.6$\,nm.
\end{abstract}

\flushbottom
\maketitle
\thispagestyle{empty}

The spin Hall effect (SHE) allows for the conversion of an electric charge current into a transverse spin current~\cite{dyakonov_possibility_1971,hirsch_spin_1999, kato_observation_2004,valenzuela_direct_2006}.
It is one of the key phenomena in the field of spintronics and has been extensively studied over the past two decades~\cite{sinova_spin_2015,manchon_current-induced_2019}.
The SHE has become an indispensable tool to generate and detect spin currents, and has been envisioned to play an important role in many spintronic devices, including next-generation magnetic memory based on spin-orbit torque (SOT).
Recent theoretical studies have indicated that in many materials the SHE arises due to the orbital Hall effect (OHE)~\cite{tanaka_intrinsic_2008,kontani_giant_2009, go_intrinsic_2018,jo_gigantic_2018,bhowal_intrinsic_2020}, arguing that the OHE is more fundamental than the  SHE~\cite{kontani_giant_2009,go_intrinsic_2018}.
The OHE converts a charge current into a transverse flow of orbital angular momentum (OAM).
It is predicted theoretically that unlike the SHE, the OHE does not require spin-orbit coupling (SOC), while in the presence of SOC, a current of orbital angular momentum can be efficiently converted into a spin current, thus converting the OHE into the SHE~\cite{tanaka_intrinsic_2008,jo_gigantic_2018}.

A few experimental studies of the orbital Hall effect were conducted so far~\cite{kim_nontrivial_2021,lee_efficient_2021,lee_orbital_2021,ding_observation_2022,sala_giant_2022,Fukunaga2023}.
They %have been 
focused on measuring SOT in non-magnetic/ferromagnetic (NM/FM) bilayer systems. 
Unexpectedly large SOT signals in seemingly trivial systems without a heavy metal, e.g.\ CuO/FM, Cr/FM, were attributed to the OHE~\cite{kim_nontrivial_2021,lee_efficient_2021,lee_orbital_2021,sala_giant_2022}. 
In such systems, however, contributions to the total measured torque acting on the FM magnetization from Rashba-type SOC at the interface~\cite{baek_spin_2018, amin_interface-generated_2018} and anomalous SOT~\cite{wang_anomalous_2019, amin_intrinsic_2019} generated by the FM itself complicate the analysis~\cite{go_theory_2020}.
It has also been recently shown that while experimentally measurable total torque is usually large and negative for Ta/FM, it depends on the choice of FM and is positive in Ta/Ni bilayers~\cite{lee_orbital_2021} as well as in Cr/Ni~\cite{lee_efficient_2021,sala_giant_2022}.
This was attributed to the torque due to OHE dominating over SHE-induced torque.
The interplay between these torques of different origin makes it difficult to disentangle a torque due to the OHE, making experimental detection of the OHE less direct.
Therefore, in order to gain an insight into the OHE it would be advantageous to study it in a single non-magnetic layer.

The magneto-optical Kerr effect (MOKE) has proved to be a valuable technique to detect spin accumulation due to the SHE in both semiconductors~\cite{kato_observation_2004} and metals~\cite{stamm_magneto-optical_2017}. 
While MOKE cannot distinguish the signals from the orbital and spin accumulation, experiments on materials with a negligible %spin-orbit coupling
SOC have a great advantage.
In these materials, the OHE can be 1-2 order of magnitude larger than the SHE, as a result, surface orbital accumulation is expected to be significantly larger than spin accumulation~\cite{jo_gigantic_2018,Salemi_theory_2022}.
Furthermore, MOKE is sensitive to the orbital magnetization because the electric dipole transition couples light helicity directly to the orbital degree of freedom, whereas the detection of spin polarization requires spin-orbit coupling. 
Therefore, signals from SHE can be reduced by choosing materials with weak SOC, e.g.\ light 3$d$ transition metals, which facilitates the separate detection of OHE as pursued recently by Choi \textit{et al.}~\cite{choi_observation_2021}, in contrast to materials with strong SOC where both OHE and SHE coexist.

In this Letter, we demonstrate the detection of the orbital Hall effect in one such material, namely, in single crystal bcc Cr(001), using longitudinal MOKE combined with ac current modulation.
MOKE microscopy with a sensitivity of about $2$\,nrad allows us to detect a current-induced magnetization at the surface of Cr which, in combination with \textit{ab initio} calculations of the OHE, SHE, and MOKE in Cr, we demonstrate to be the orbital accumulation due to the OHE.
Comparison of the experimental data with the theoretical calculations yields quantitative values for the orbital Hall angle and the orbital diffusion length of Cr.
Thus, our results demonstrate a possibility to detect and quantify the OHE in a NM, independent of an adjacent FM.
Along with Choi \textit{et al.}~\cite{choi_observation_2021}, this provides strong evidence for the OHE.

Experiments are performed on Cr(001) films grown by molecular beam epitaxy on MgO(001) substrates with film thicknesses ranging between 7.5 and 50\,nm.
Details on the growth and characterization of Cr films is given in Supplemental Material, Sec.~S1~ (SM:S1)~\cite{SupplMat}.
All Cr films are capped with 5\,nm CaF$_2$ to protect them from oxidation, unless otherwise indicated.
The samples are lithographically patterned into $20\,\mu$m wide channels for the MOKE measurements and Hall bars for resistivity measurements.

In the longitudinal MOKE measurements, a \textit{s}-polarized laser beam (to exclude contribution from transverse MOKE) with wavelength $\lambda = 800$\,nm is incident at $45\degree$ from the sample normal.
The laser beam is focused to a $\sim$3\,$\mu$m spot size on the sample using an objective lens.
The reflected light is collected using a second objective positioned at $90\degree$ to the first one.
The sample is mounted on an \textit{xy} translation stage which allows scanning across the sample surface.
A sine-modulated current is applied to the lithographically patterned device (patterned along Cr [100] direction), inducing orbital accumulation on the top and bottom surfaces. 
The Kerr rotation of the reflected light is measured using a combination of a half-wave plate, a polarization-splitting Wollaston prism, a balanced photodetector, and a lock-in amplifier at the frequency of the current modulation (details in SM:S2,S3).

\begin{figure}
    \subfloat{\xincludegraphics[scale=0.027,label=(a)]{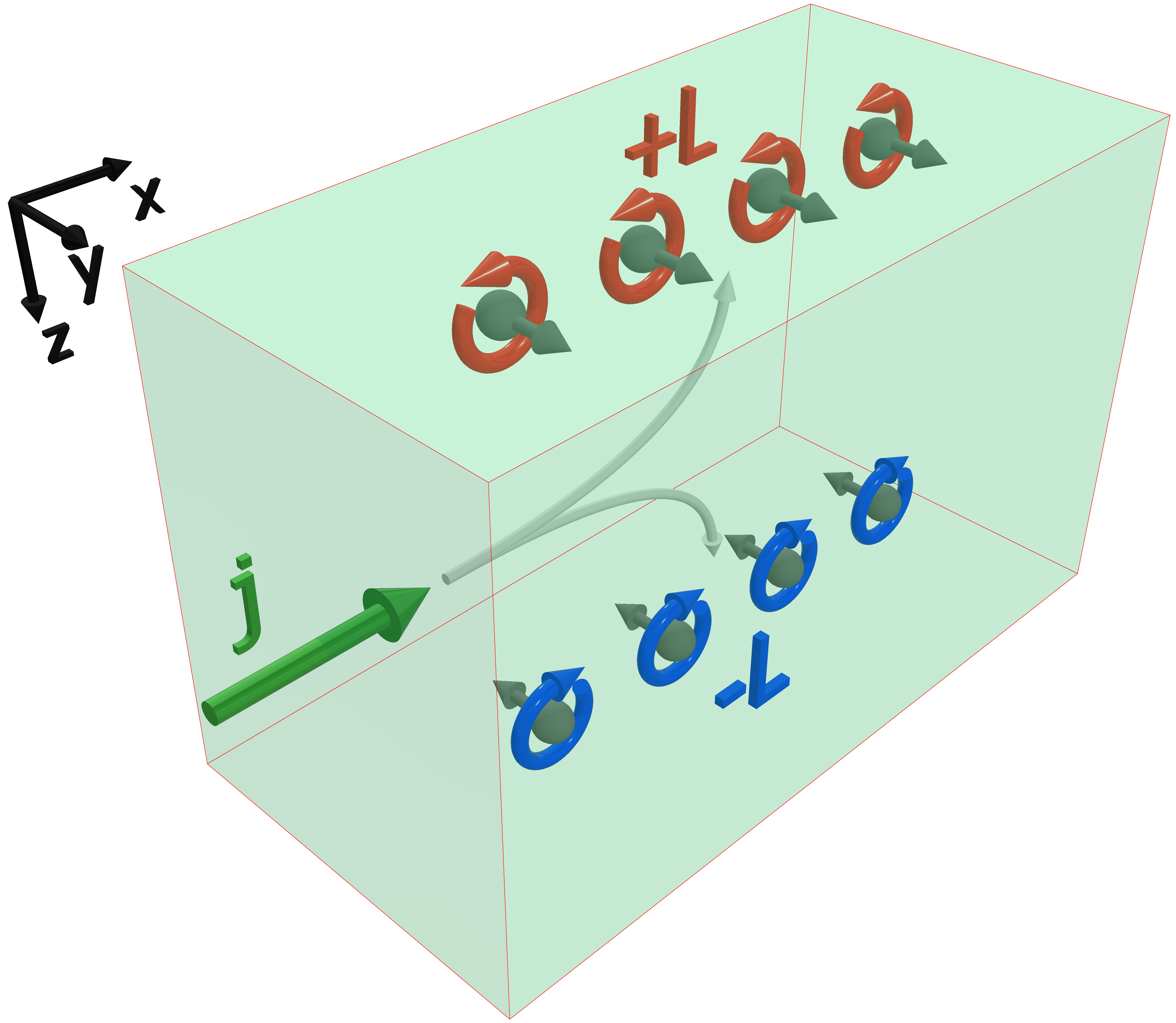}\label{fig:schematics_of_OHE}}
  \hfill
   \subfloat{\xincludegraphics[scale=0.525,label=(b)]{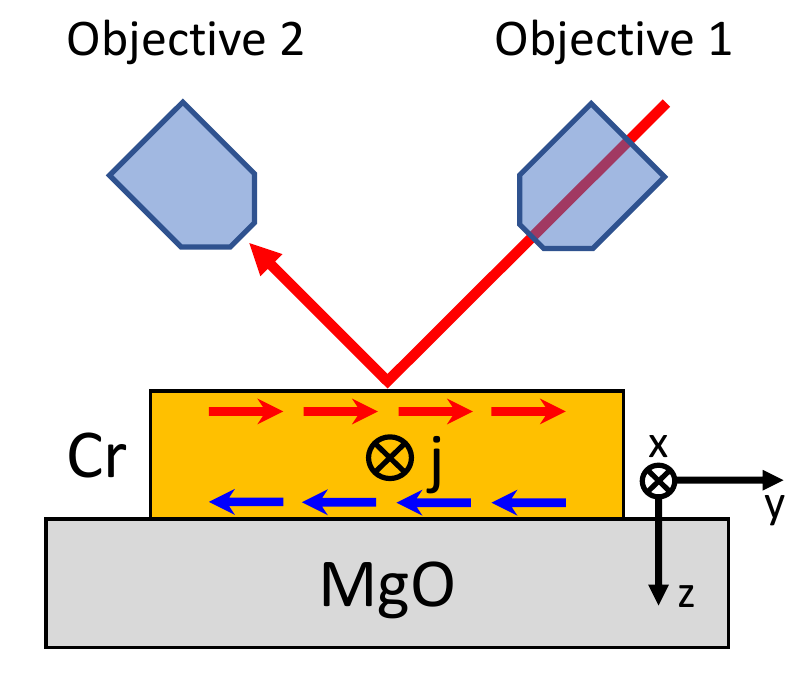}\label{fig:schematics_of_experiment}}
%  \hfill
   % \subfloat{\xincludegraphics[scale=1,label=c)]{Orbital accumulation vs t.pdf}\label{fig:Orbital_vs_thickness}}
  \caption{(a) Schematics of the orbital Hall effect. The charge current $j$ generates a transverse orbital current, leading to orbital accumulation on the sample's surfaces. (b) Measurement setup utilizing the longitudinal MOKE to detect the in-plane orbital accumulation.
  }
\end{figure}

Figure~\ref{fig:schematics_of_experiment} shows the longitudinal MOKE measurement geometry.
The OHE generates an orbital accumulation oriented along the in-plane \textit{y} direction transverse to the electric current flowing along \textit{x}.
The generated orbital accumulation has opposite signs at the opposite surfaces of the Cr film.
The longitudinal MOKE detects the superposition of the polarization rotation caused by the \textit{y}-polarized OAM accumulated at the top and bottom surfaces.
For quantitative analysis, one needs to take into account the
light attenuation in a metallic film, with the characteristic penetration length of the light usually on the order of 10-30\,nm.
At the same time, the orbital diffusion length determines the spatial distribution and the amount of OAM accumulation, which is reduced for films thinner than $l_o$.
For sufficiently thick films these two effects will lead to a saturation of the OAM accumulation detected by MOKE.
Therefore, a thickness-dependent study can estimate orbital diffusion length in a single layer film.
This is important as the characteristic length of orbital transport is debated theoretically, with some works predicting very short orbital accumulation  distances on the order of several atomic layers~\cite{salemi_quantitative_2021}, and other works predicting orbital diffusion distances longer than the spin diffusion length~\cite{choi_observation_2021,Hayashi2023,go_long-range_2021}.

\begin{figure}
    \subfloat{\xincludegraphics[width=0.235\textwidth,label=(a)]{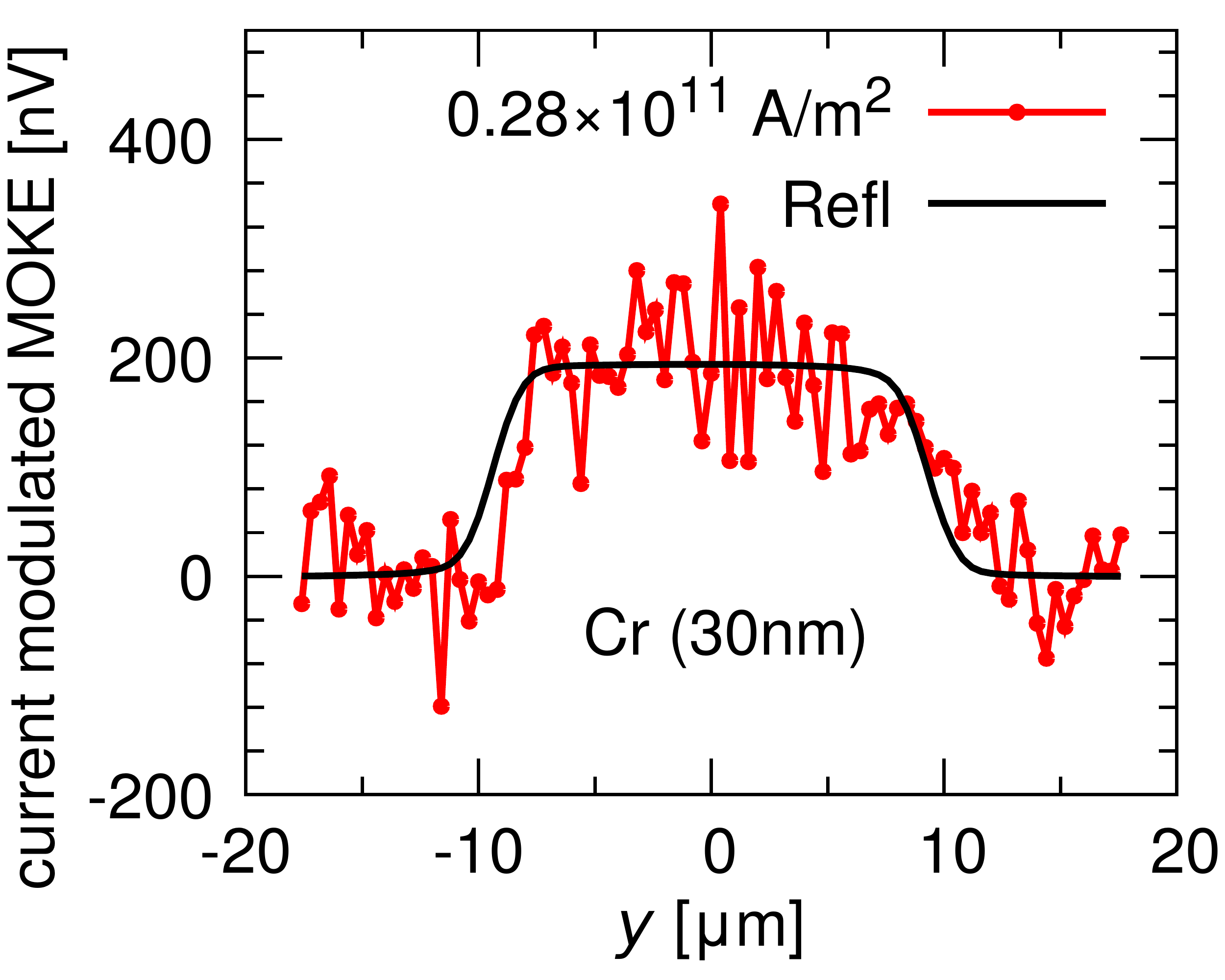}\label{fig:Cr_30nm_linecut}}
  \hfill
    \subfloat{\xincludegraphics[width=0.235\textwidth,label=(b)]{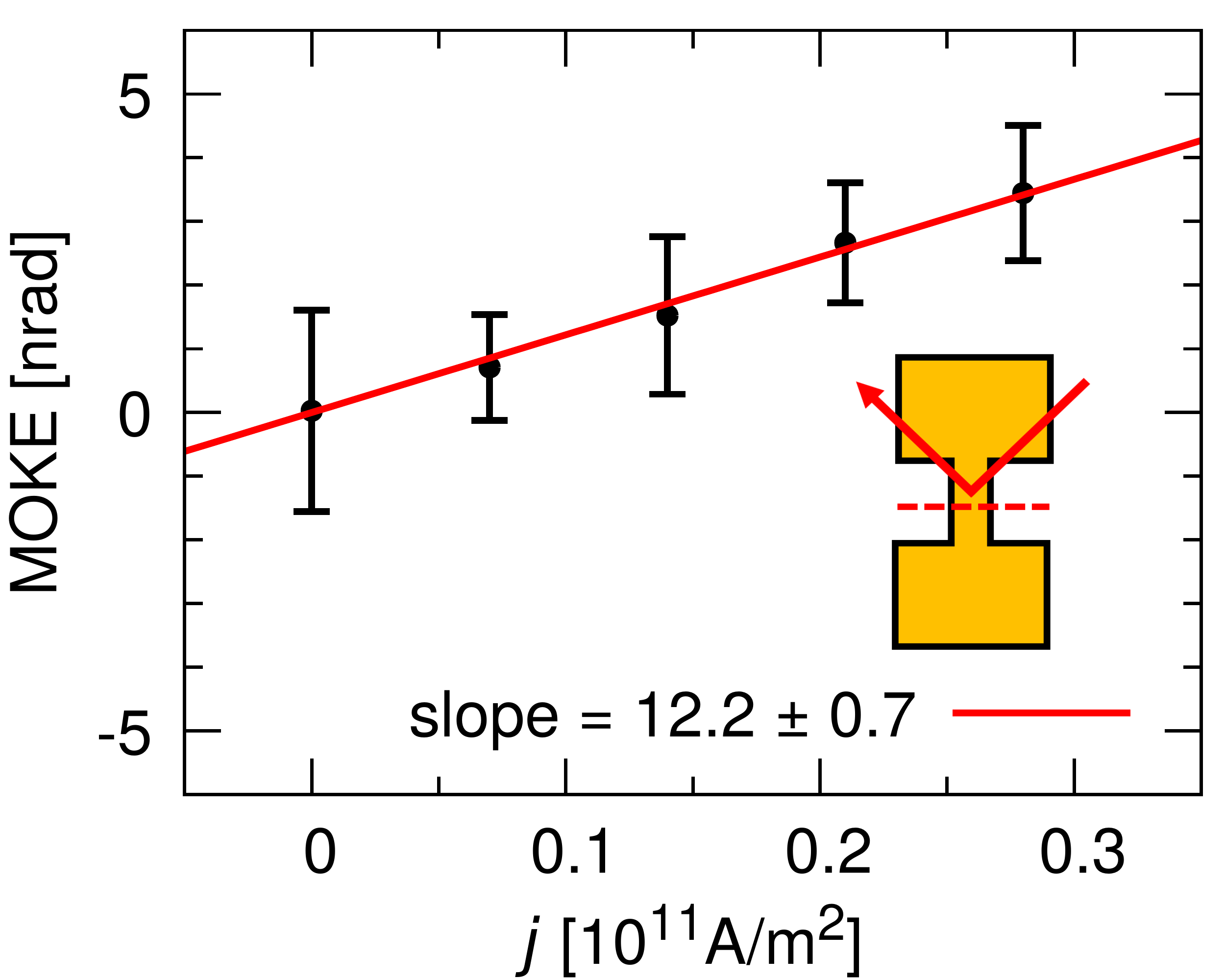}\label{fig:Cr_moke_vs_current_vert}}
  \hfill
    \subfloat{\xincludegraphics[width=0.235\textwidth,label=(c)]{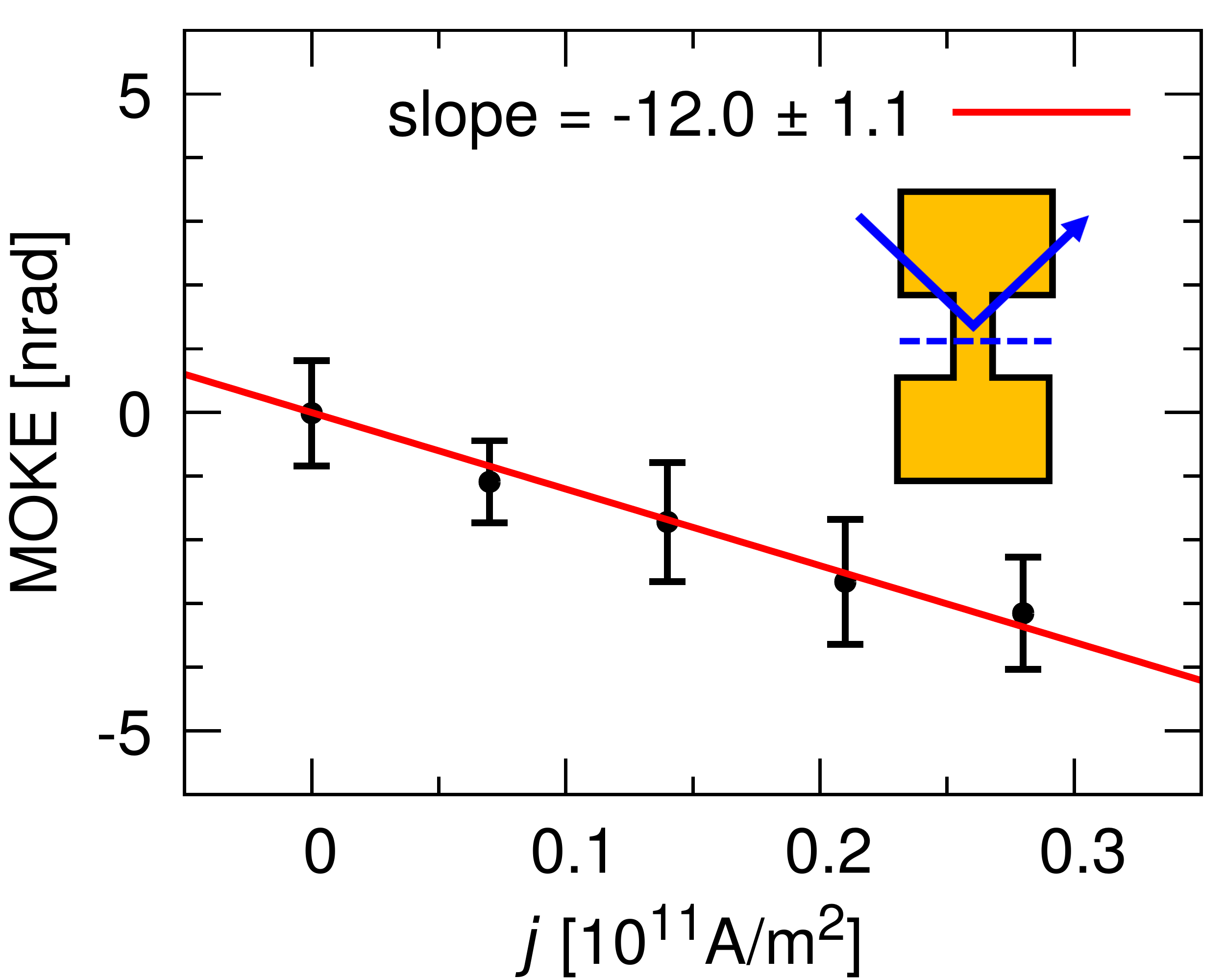}\label{fig:Cr_moke_vs_current_vert_rev_path}}
  \hfill
    \subfloat{\xincludegraphics[width=0.235\textwidth,label=(d)]{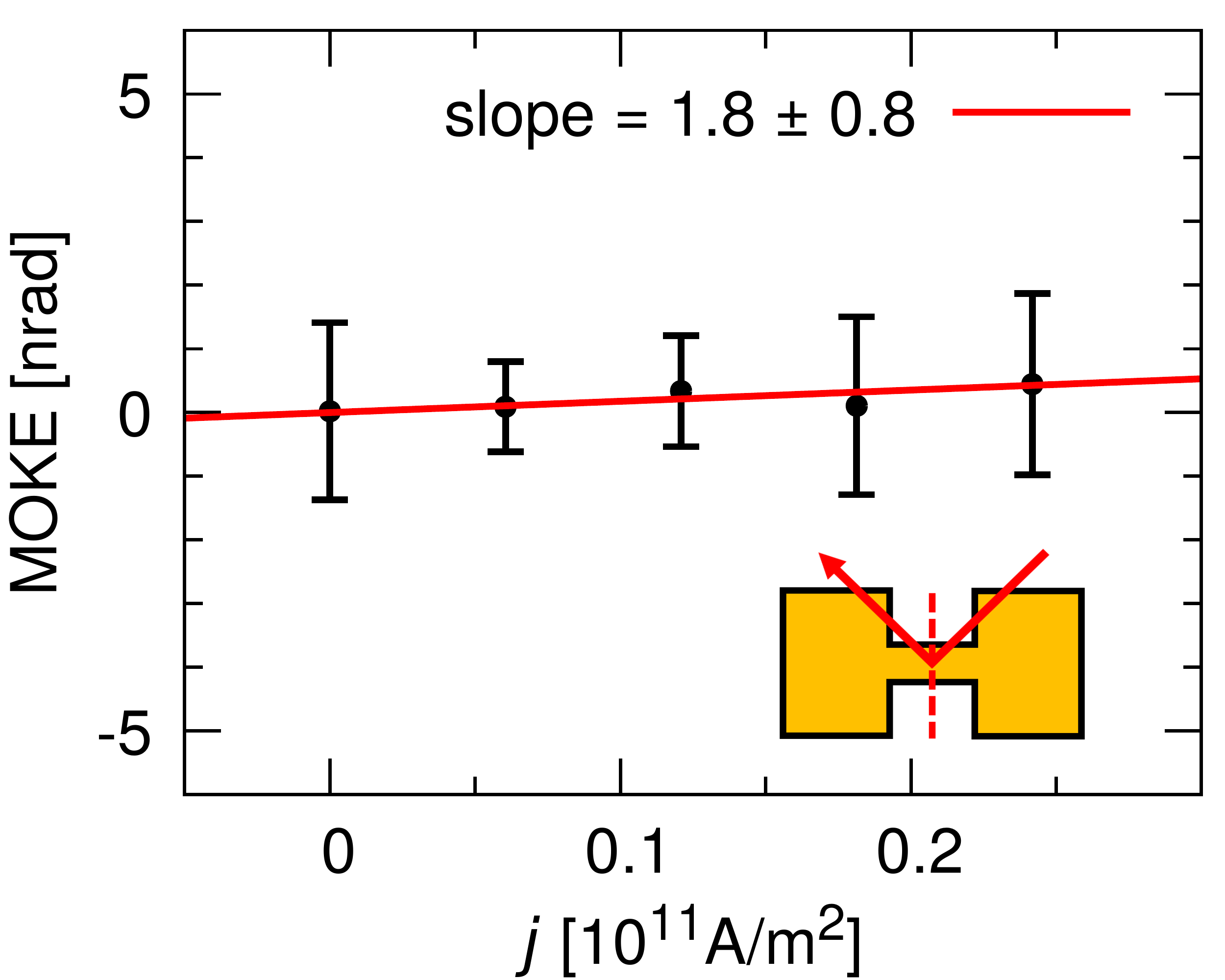}\label{fig:Cr_moke_vs_current_horz}}
    \vspace*{-0.2cm}
  \caption{Magneto-optical Kerr rotation induced by the orbital and spin Hall effect.
  (a) MOKE line scan across a 20\,$\mu$m wide, 30\,nm thick Cr channel at rms current density $j=0.28 \times 10^{11}$\,A/m$^2$.
  MOKE signal as a function of current density: 
  (b) for the laser beam incidence shown in Fig.~\ref{fig:schematics_of_experiment},
  (c) for the reversed optical path of the laser beam (30\,nm sample),
  (d) for device rotated $90\degree$ about the surface normal (40\,nm sample).
  }
  \label{fig:MOKE_Cr_main}
\end{figure}

Stamm \textit{et al.}~\cite{stamm_magneto-optical_2017} has shown that special care should be taken to detect MOKE signals due to the SHE in Pt and W in order to avoid artificial signals due to a change in reflectivity of the sample caused by Joule heating $\sim j^2$ (SM:S4). 
Similar to Stamm \textit{et al.}~\cite{stamm_magneto-optical_2017}, 
we measure the current-induced Kerr rotation as the voltage output of the balanced photodetector at the frequency of the sinusoidal driving current and subtraction of MOKE signals measured with opposite current polarities (inverted by changing the wiring of the device) is performed to exclude any remaining thermal artifacts.
Figure~\ref{fig:Cr_30nm_linecut} shows the Kerr rotation measured on a 30\,nm thick Cr film at current density of $0.26 \times 10^{11}$\,A/m$^2$.
We observe a Kerr rotation signal from the surface of
the conducting Cr channel, which is of the order of a few nanoradians. 
The MOKE signal is approximately constant over the Cr channel surface (SM:S5) and scales linearly with the applied current, consistent with the orbital/spin accumulation picture.
The absence of antisymmetry in the \textit{y} spatial profile in Fig.~\ref{fig:Cr_30nm_linecut} as well as measurements on a Cr/MgO/Cr sample show that the MOKE signal does not originate from an in-plane Oersted field (see SM:S6).
The value of Kerr rotation per current density for a 30\,nm Cr sample is $12.2 \pm 0.7$\,nrad per $10^{11}$\,A/m$^2$ as found from a linear fit to the data shown in Fig.~\ref{fig:Cr_moke_vs_current_vert}.

Further evidence that the measured Kerr rotation is consistent with the \textit{y}-polarized orbital accumulation in Cr comes from the following control measurements.

\begin{figure}
    \subfloat{\xincludegraphics[width=0.235\textwidth,label=(a)]{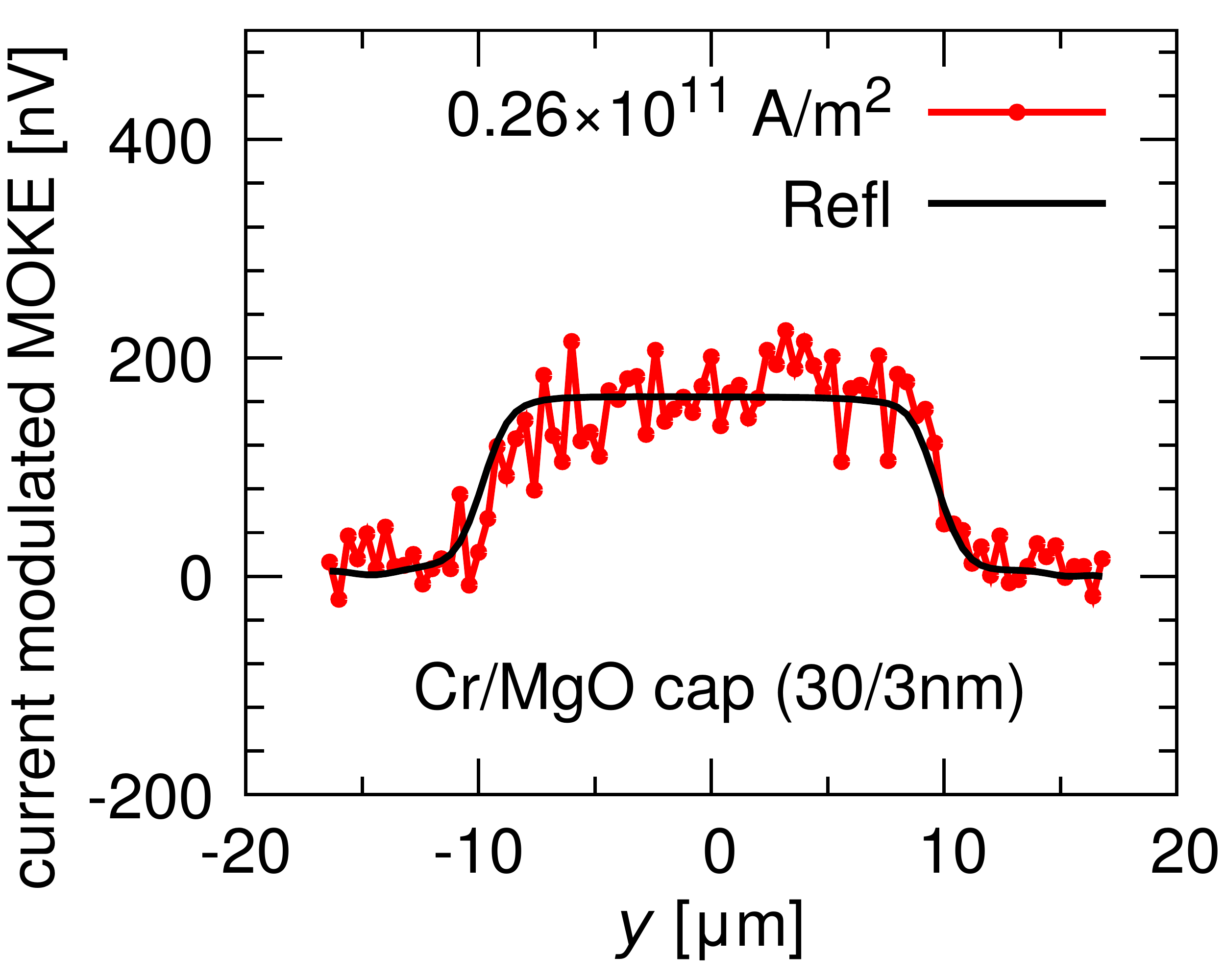}\label{fig:Cr_MgO_linecut}}
  \hfill
    \subfloat{\xincludegraphics[width=0.235\textwidth,label=(b)]{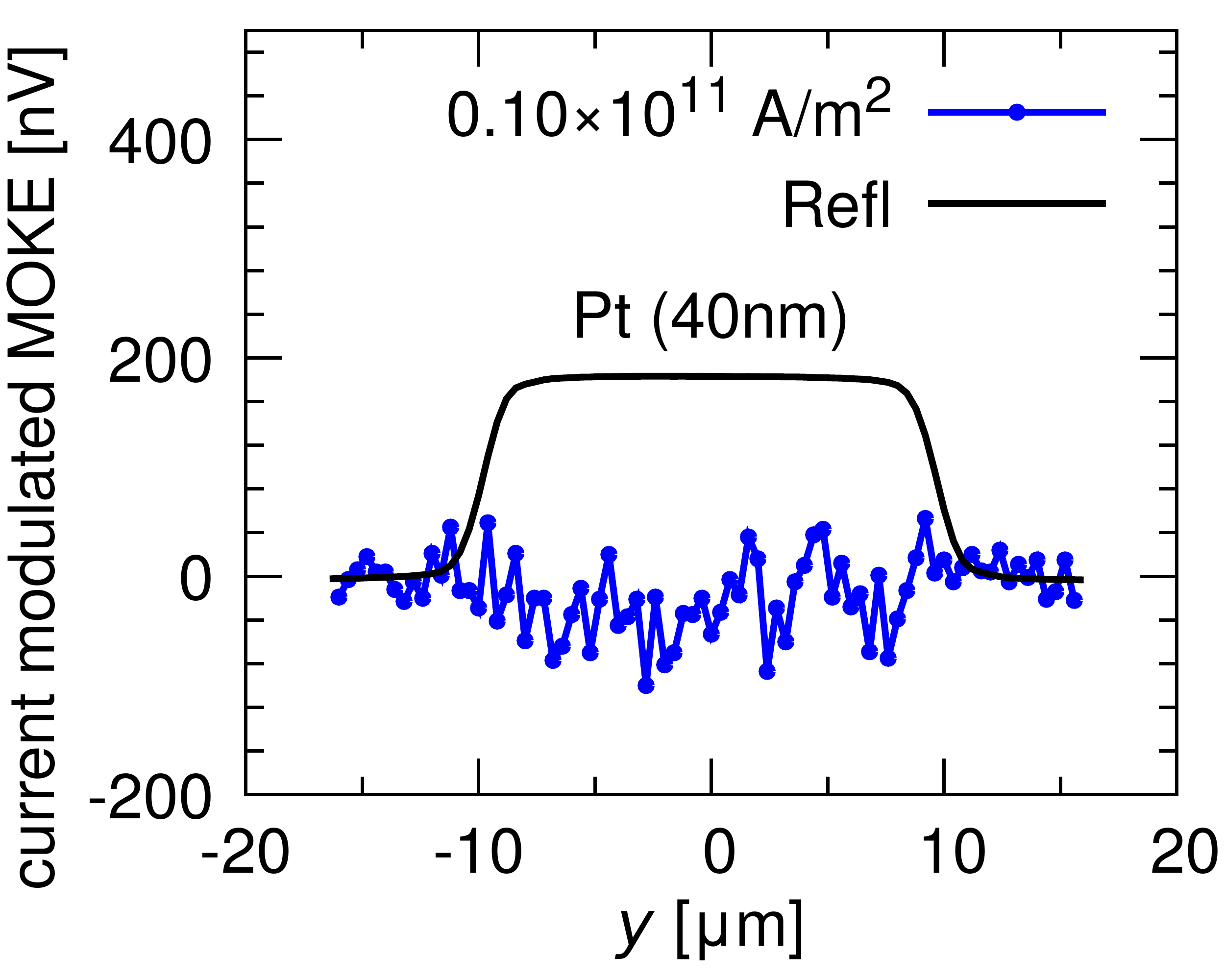}\label{fig:Pt_40nm_linecut}}
  \hfill
    \subfloat{\xincludegraphics[width=0.235\textwidth,label=(c)]{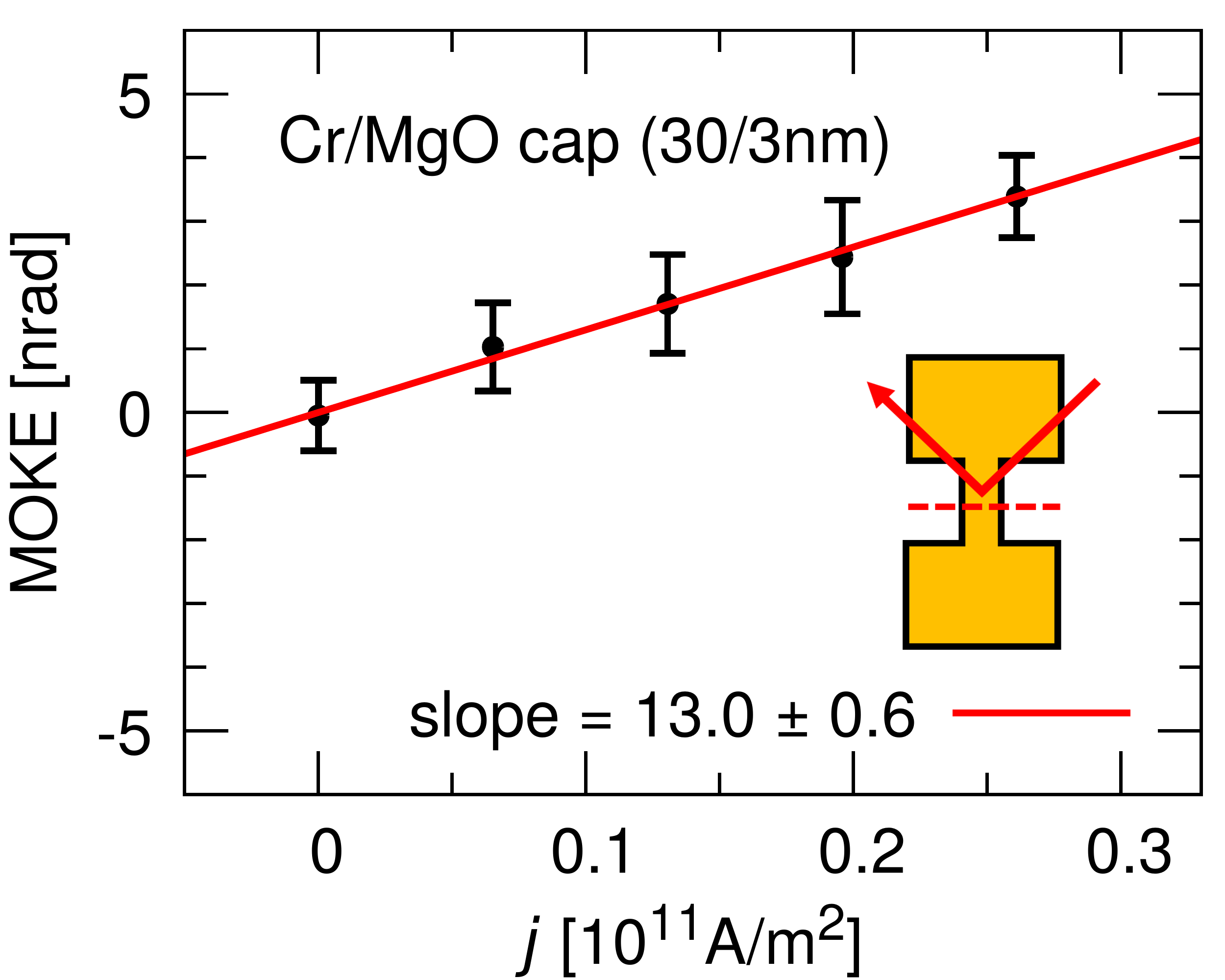}\label{fig:Cr_MgO_moke_vs_current_vert}}
  \hfill
    \subfloat{\xincludegraphics[width=0.235\textwidth,label=(d)]{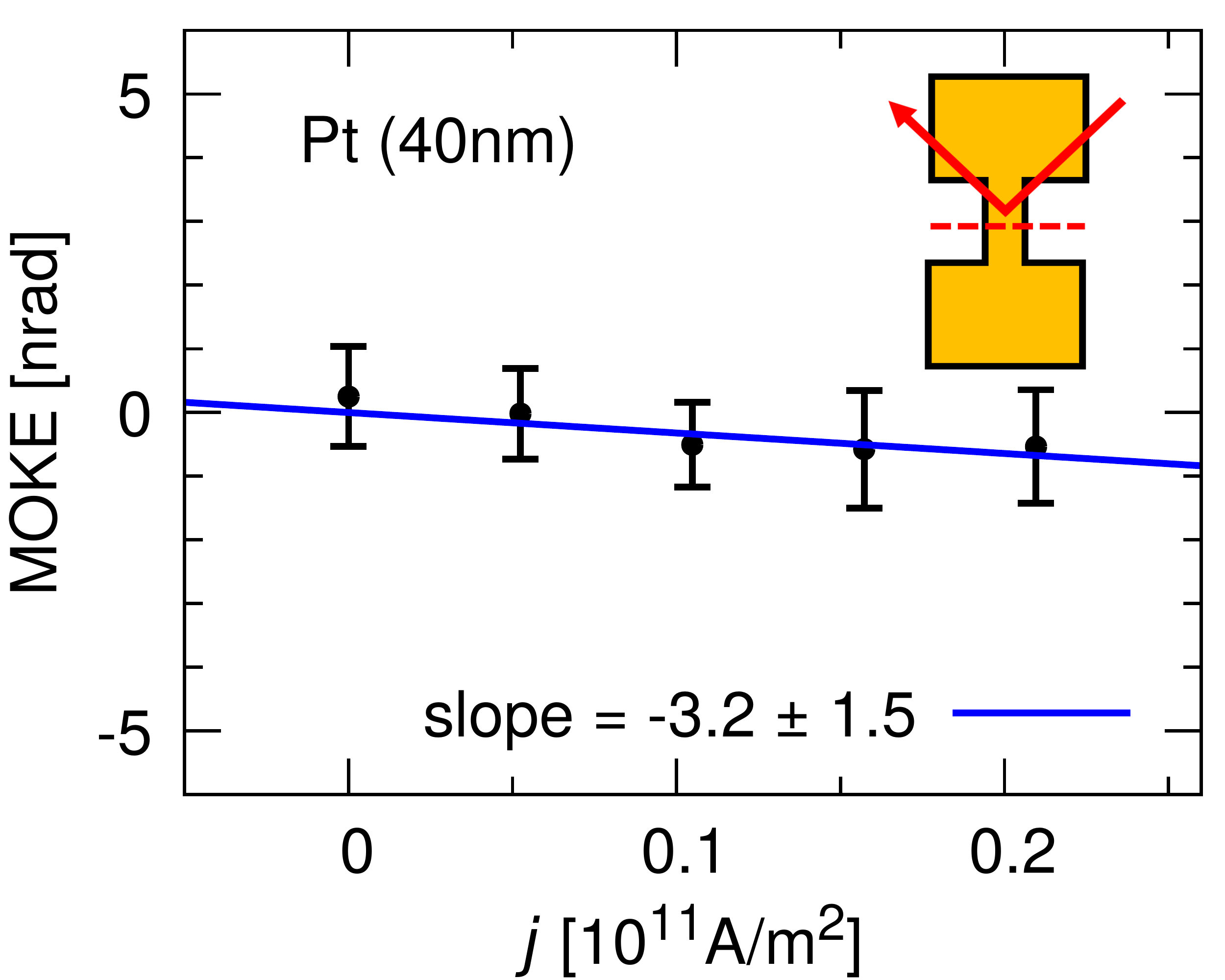}\label{fig:Pt_moke_vs_current_vert}}
    \vspace*{-0.2cm}
  \caption{Magneto-optical Kerr rotation induced by the orbital Hall effect.
    (a) MOKE line scan across a 20\,$\mu$m wide Cr/MgO (30/3\,nm) channel at current density $j=0.26 \times 10^{11}$\,A/m$^2$
    (b) MOKE line scan across a 40\,nm thick Pt channel at $j=0.10 \times 10^{11}$\,A/m$^2$
    (c) and (d) MOKE signal as a function of current density for Cr and Pt, respectively. }
  \label{fig:MOKE_Cr_and_Pt}
\end{figure}

First, the longitudinal MOKE geometry chosen in our work is sensitive to in-plane \textit{y} and out-of-plane \textit{z} components of magnetization.
Upon inversion of the light incidence angle, the Kerr rotation due to the \textit{y} component should change sign,
%to the opposite, 
while the Kerr rotation due to the \textit{z} component should stay constant.
Figure~\ref{fig:Cr_moke_vs_current_vert_rev_path} shows that the Kerr rotation measured with the opposite angles of incidence reverses sign. 
The linear fits to the data in Fig.~\ref{fig:Cr_moke_vs_current_vert} and~\ref{fig:Cr_moke_vs_current_vert_rev_path} yield %the 
slopes that, within error bars, are the same.
The linear dependence with current density demonstrates the absence of a thermally induced signal and the same magnitudes of the slopes exclude the presence of a polar MOKE contribution due to a magnetization along \textit{z}.

Second, if the measured MOKE signal originates from an orbital accumulation due to the OHE, then when the sample is rotated by $90\degree$ about the normal, the Kerr rotation should decrease to zero, as the orbital accumulation is polarized along \textit{x} in this case.
Figure~\ref{fig:Cr_moke_vs_current_horz} shows that upon rotating sample by $90\degree$ Kerr signal indeed decreases and can be said to be zero within error bars (SM:S7). 
This behavior is consistent with the OHE/SHE origin of the signal.

Third, to investigate whether the observed signals are sensitive to the interface, we insert 3\,nm of MgO in between the Cr and CaF$_2$ capping layer.
Fig.~\ref{fig:Cr_MgO_moke_vs_current_vert} shows that the different interface does not change the magnitude of the signal.
The measured Kerr rotation is the same, within error bars, for the devices with Cr/MgO and Cr/CaF$_2$ interfaces.
This result demonstrates that it is unlikely that the origin of the effect is interfacial (e.g., spin or orbital Rashba-Edelstein effect), and points to the bulk origin of the effect, as expected theoretically for OHE.

Fourth, we compare the signs of Kerr rotation in Cr and Pt films.
Fig.~\ref{fig:Pt_40nm_linecut} and \ref{fig:Pt_moke_vs_current_vert} show the experimentally measured MOKE signal for a sputtered 40\,nm Pt sample. 
The value of Kerr rotation per current density has a value smaller in magnitude than measured in Cr and of the opposite sign, $-3.2\pm1.5$\,nrad per $10^{11}$\,A/m$^2$.
We find that the opposite signs in Cr and Pt are consistent with the opposite Kerr coefficients theoretically calculated for Cr and Pt at 800\,nm wavelength (SM:S11).
To follow the sign convention in Stamm \textit{et al.}~\cite{stamm_magneto-optical_2017}, we choose the sign of current-induced Kerr rotation in Pt to be negative (see SM:S8).
To confirm the magnitude and sign of MOKE in Pt, we conducted measurements on an MBE-grown Pt sample which, within error bars, yields the same MOKE values (SM:S9).

\begin{figure}
%\hspace*{-0.5cm}
\includegraphics[width=0.99\linewidth]{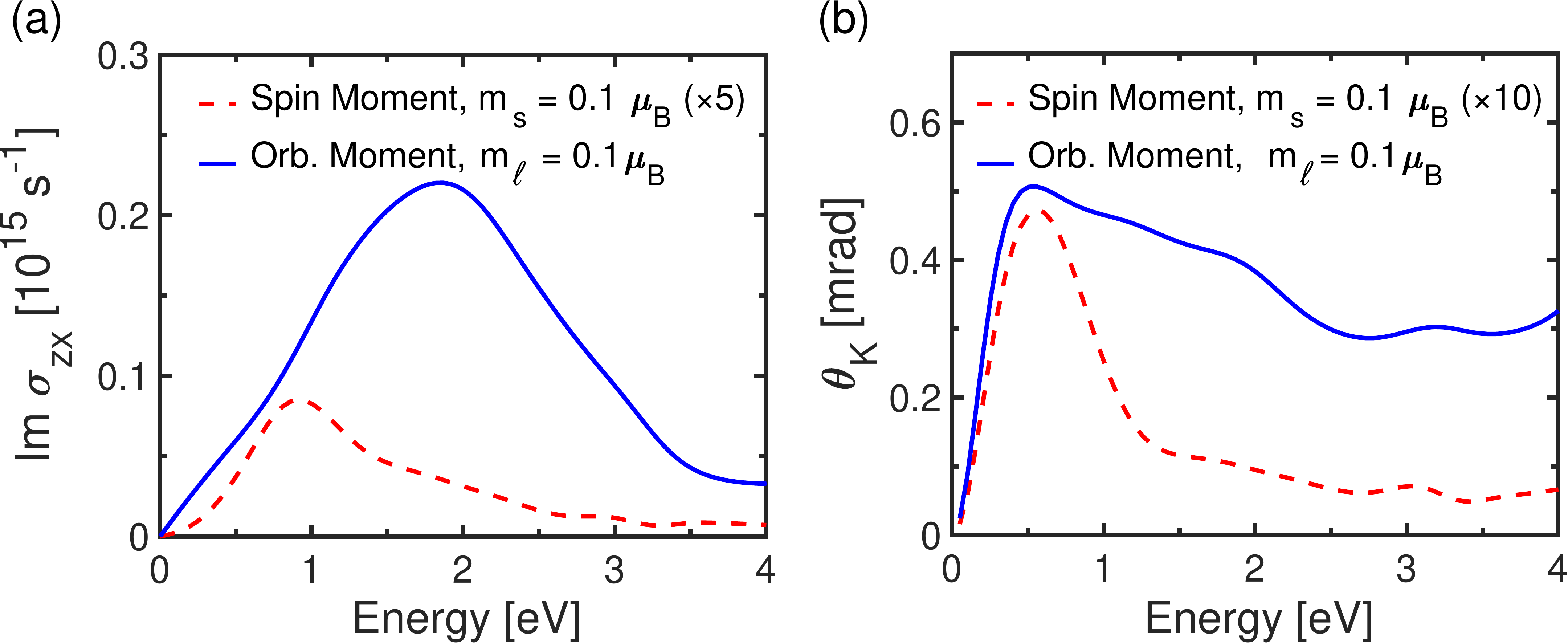}  
  \vspace*{-0.5cm}
%\subfloat{\xincludegraphics[width=0.2360\textwidth,label=(a)]{figures/Im_zx.eps}\label{fig:ab_initio_Imxy_Cr}}
%    \vspace*{-0.6cm}
  \hfill
  \caption{(a) The \textit{ab initio} calculated magneto-optical conductivity $\textrm{Im}\,\sigma_{zx}(\omega)$ of Cr for an induced spin moment $m_s$ or orbital moment $m_{\ell}$ of 0.1\,$\mu_{\rm B}$. (b) The calculated longitudinal Kerr rotation $\theta_{\rm K}$ of Cr due to a spin or an orbital moment. Note that $\theta_{\rm K}$ for the induced spin polarization is multiplied by 10.}
  \label{fig:ab_initio}
\end{figure}

Fifth, to examine the orbital diffusion in Cr we perform \textit{ab initio} calculations of the OHE, SHE, and wavelength-dependent longitudinal MOKE.
To compute the magneto-optical signal due to a spin or an orbital polarization in Cr metal, we model the induced spin and orbital moment by including a Zeeman term in the relativistic DFT Hamiltonian, $\mathcal{H}= \mu_{\rm B} \bm{B}_{ext} \cdot (\bm{\ell} +2 \bm{s})$, with $\bm{B}_{ext}$ the applied field and $\bm{\ell}$ and $\bm{s}$ the orbital and spin angular momentum operators, respectively. 
Coupling $\bm{B}_{ext}$ selectively to \textit{either} spin or orbital angular momentum gives us an induced spin or orbital polarization.
We then use the linear-response formalism to compute the magneto-optical spectrum \cite{Oppeneer2001} for either a spin or an orbital polarization in bulk Cr (SM:S10). 
Note that when we couple e.g.\ the Zeeman field only to the spin, there will nonetheless be a very small orbital moment generated by the nonzero spin-orbit coupling. We find that the magneto-optical conductivity can be approximated as
$\sigma_{zx} (\omega ) = \sigma_{zx}^{s} (\omega ) m_s + \sigma_{zx}^{o} (\omega ) m_{\ell}$, i.e., there are two distinct contributions that  are linear in the spin or orbital moment. 
In Fig.\ \ref{fig:ab_initio}(a) we show the \textit{ab initio} calculated magneto-optical conductivity spectra for induced spin or orbital polarizations. 
It can be recognized immediately that an induced orbital moment leads to a much larger $\textrm{Im}\,\sigma_{zx} (\omega )$ than an induced spin moment. 
Next, we use the calculated optical conductivity $\sigma_{xx} (\omega )$ and compute the resulting longitudinal MOKE for a $45^{\circ}$-angle of incidence (see \cite{Oppeneer2001} and SM:S10 for details). 
The calculated Kerr rotation spectra for bulk Cr are shown in Fig.\ \ref{fig:ab_initio}(b). 
Again, we observe that the Kerr rotation generated by orbital polarization is substantially larger ($> 10\times$) than that due to spin polarization.

\begin{figure}
\includegraphics[width=0.38\textwidth]{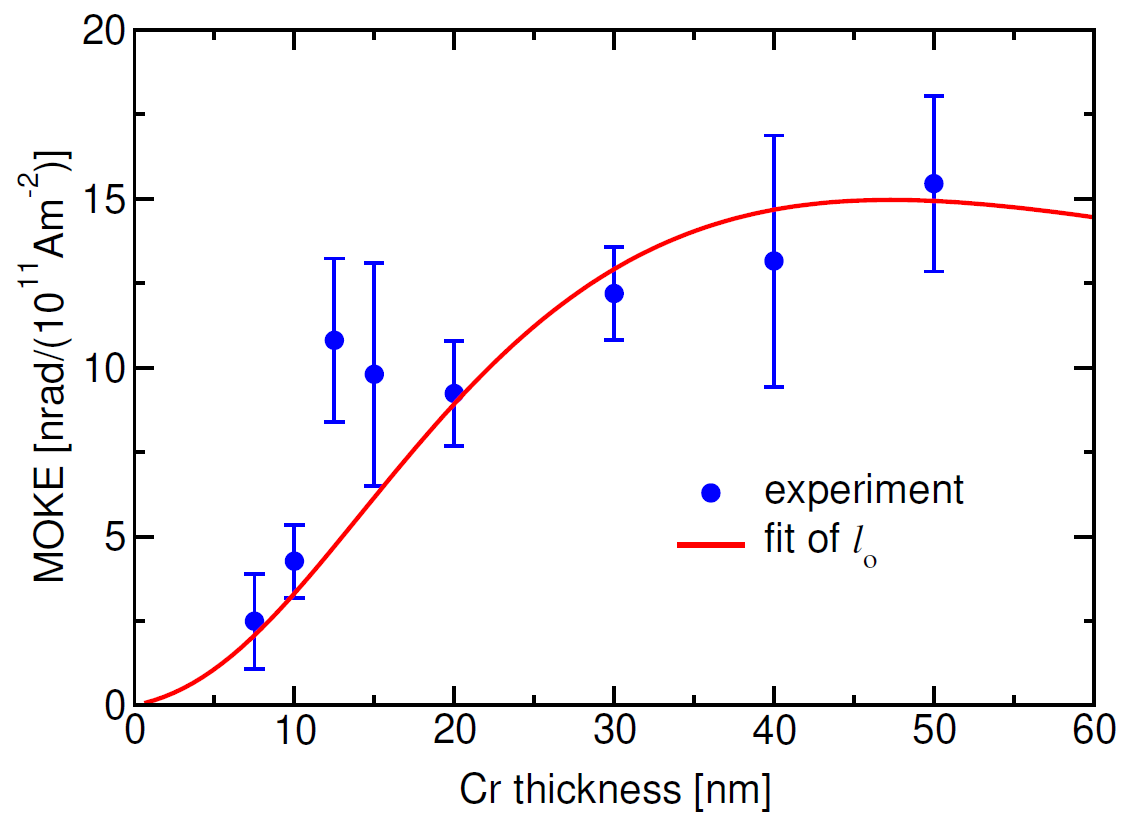}
\vspace*{-0.3cm}
  \caption{Measured Kerr rotation $\theta_{\rm K}$ as function of Cr film thickness (symbols), for a current density of $10^{11}$ Am$^{-2}$. The full curve is a fit to the measured MOKE rotation based on Eq.\ (\ref{eq:meask}) with \textit{ab initio} calculated quantities to obtain the orbital diffusion length $l_o$.}
  \label{fig:thickness_dependence}
\end{figure}

After verifying that detected MOKE signal in Cr (001) is not a thermal artifact and fully consistent with the symmetries of the OHE, we investigate the characteristic length of the orbital transport by studying the Kerr rotation as a function of Cr film thickness, $t$.
Figure~\ref{fig:thickness_dependence} shows the experimental results for the
measured Kerr rotation normalized by a current density $j = 10^{11}$\,A/m$^{2}$.
Intuitively, three effects can contribute to the thickness dependence: the saturation of orbital accumulation with a length scale of $l_o$ as discussed earlier, the limited penetration depth of the light ($\sim 18$\,nm in Cr, see SM:S10), and the opposite sign of the orbital accumulation at the top and bottom interfaces.

To model the thickness-dependent MOKE we use the formalism of Ref.\ \cite{stamm_magneto-optical_2017}, but suitably modified to describe separately the Kerr rotations due to orbital and spin accumulation, respectively. Thus, we assume   that orbital diffusion is described by expressions equivalent to those for spin diffusion. The thickness-dependent Kerr rotation due to orbital accumulation is given by
\begin{eqnarray}
\label{eq:meask}
\!\! \!\!\theta_{\rm K}^{o} &=& \frac{l_o \sigma_{zx}^{\rm OH} \, \rho(t)^2 j D(E_{\rm F}) \, \mathrm{e}^{\frac{t}{2 l_o}}}{\cosh(\frac{t}{2l_o})} \times \nonumber \\
& {\rm Re}& \left\{ \Phi_{\rm K}^{bulk,o}\kappa \bigg( \frac{(\mathrm{e}^{-\kappa^-t}-1)\mathrm{e}^{-\frac{t}{l_o}}}{\kappa^-}-\frac{\mathrm{e}^{-\kappa^+t}-1}{\kappa^+}\bigg)
\! \right\} \! ,
\end{eqnarray}
where $\sigma_{zx}^{\rm OH}$ is the bulk OHE, $\rho(t)$ the resistivity, $D(E_{\rm F})$ the density of states at the Fermi energy $E_{\rm F}$, $\Phi_{\rm K}^{bulk,o}$ is the bulk complex Kerr effect due to orbital polarization, and $\kappa= (4\pi i \bar{n} \cos\psi )/\lambda $, with wavelength $\lambda$, $\cos \psi= (1-\sin^2\phi_i/\bar{n}^2)^{1/2}$, with $\bar{n}$ the complex index of refraction, $\phi_i$ the angle of incidence, and $\kappa^{\pm}=\kappa \pm 1/l_o$. A similar expression exists for $\theta_{\rm K}^s$ \cite{stamm_magneto-optical_2017}, but containing the spin diffusion length $l_s$, the SHE conductivity $\sigma_{zx}^{\rm SH}$, and bulk Kerr effect due to spin polarization, $\Phi_{\rm K}^{bulk,s}$.

Our \textit{ab initio} calculations provide values $\sigma_{zx}^{\rm SH}$ and $\sigma_{zx}^{\rm OH}$ as well as $\Phi_{\rm K}^{bulk,s}$ and $\Phi_{\rm K}^{bulk,o}$. $\bar{n}$ is evaluated from the computed optical conductivity $\sigma_{xx}(\omega)$, which is in good agreement with measurements (SM:S10). 
Using both \textit{ab initio} calculated quantities and  quantities that are given from the experiment ($\rho(t)$, $t$, $j$, $\phi_i$) we can use the measured MOKE as function of thickness together with Eq.\ (\ref{eq:meask}) to obtain values for $l_o$ and $l_s$.
At this point it is instructive to mention that the calculated \cite{Salemi_theory_2022}
$\sigma_{zx}^{\rm OH}$ is $ 7000\, (\hbar/e)$ $\Omega^{-1}\rm{cm}^{-1}$ whereas $\sigma_{zx}^{\rm SH} \approx -70\, (\hbar/e)$ $\Omega^{-1}\rm{cm}^{-1}$, i.e., much smaller. 
As a consequence, we find that the measured thickness-dependent $\theta_{\rm K}$ data in Fig.~\ref{fig:thickness_dependence} cannot be described by spin diffusion. 
This provides our fifth evidence that the observed MOKE signal is due to orbital accumulation.

To extract the orbital diffusion length from the measured data, we adopt two approaches. First, we can divide $\theta_{\rm K}^o$ in Eq.\ (\ref{eq:meask}) by the measured resistivity $\rho(t)^2$ (SM:S12).
The data for $\theta_{\rm K}^o/\rho^2$ as function of $t$ can then be fitted by using the theoretical input values to obtain $l_o$.
Second, we can model the thickness-dependent resistivity by first fitting the resistivity data $\rho(t)$ with the Fuchs-Sondheimer equation \cite{Fuchs1938,Sondheimer1952} (SM:S13).
After that, we use the obtained resistivities as input values in Eq.\ (\ref{eq:meask}) and fit $\theta_{\rm K}^o$ as a function of $t$ to obtain $l_o$.
In Fig.\ \ref{fig:thickness_dependence} we show the result from the fit of $\theta_{\rm K}^o$ with Eq.\ (\ref{eq:meask}) (second procedure) together with the experimental data.
The theoretical curve based on \textit{ab initio} calculated values for $\sigma_{zx}^{\rm OH}$, $\kappa$, and $\Phi_{\rm K}^{bulk,o}$ captures the measured MOKE data very well.
The fit value of the orbital diffusion length is $l_o = 6.66 \pm 0.48$ nm.
Applying the first procedure, we obtain $l_o = 6.48 \pm 0.62$ nm (SM:S14).
Both approaches give thus consistently very similar values, from which we find an averaged orbital diffusion length $l_o = 6.6\pm 0.6$ nm.

Using the theoretically calculated value of the orbital Hall conductivity and the average resistivity of the films (SM:S12), we can estimate an orbital Hall angle in our Cr films, $\theta^{\rm OH} = \sigma_{zx}^{\rm OH} \rho\,(2e/\hbar) \approx 0.28$.
This value is comparable to values of the effective spin Hall angle in Pt/FM bilayers~\cite{manchon_current-induced_2019,zhu_maximizing_2021}.

Orbital transport is currently only poorly understood.
A key unsolved question is the size of the orbital diffusion length.
So far, a rather wide range of orbital diffusion lengths (from 3\,nm  to 74\,nm) have been reported \cite{lee_efficient_2021,choi_observation_2021,Hayashi2023,Liu2023,Fukunaga2023}.  
Liu \textit{et al.}\ \cite{Liu2023} obtained $l_o \sim 3.1$\,nm for Nb from torque measurements on Co/Nb bilayers.
Fukunaga \textit{et al.}\ \cite{Fukunaga2023} extracted from torque measurements on Ni/Zr an $l_o \approx 9.7$\,nm for Zr. 
%These values are comparable to our $l_o$ for Cr.
Using MOKE as well as torque measurements, Choi \textit{et al.}\ \cite{choi_observation_2021} determined an orbital diffusion length of $61 - 74$\,nm for Ti.
However, they had to reduce the \textit{ab initio} computed $\sigma^{\rm OH}$ by a factor of hundred to match it with their MOKE measurements.
From torque measurements on a Ni/Ti bilayer, Hayashi \textit{et al.}\ \cite{Hayashi2023} deduced $l_o \approx 50$\,nm for Ti.
Both these values are roughly ten times larger than our $l_o$ for Cr.
Interestingly, Lee \textit{et al.}\ \cite{lee_efficient_2021} performed torque measurements on Ni/Cr bilayers and obtained $l_o = 6.1 \pm 1.7$\,nm for Cr. This is remarkably consistent with the $l_o$ value we obtained using magneto-optics on a single Cr layer, which arguably should provide more accurate, intrinsic $l_o$ values.
It should be emphasized that the orbital diffusion length of Cr is longer than the Cr spin diffusion length of 4.5\,nm, determined on Fe/Cr films \cite{Bass2007}. 
Consequently, long-range orbital transport carried by a huge orbital conductivity is expected for Cr.

In conclusion, we have used MOKE microscopy combined with {\textit{ab initio}} calculations of MOKE and orbital Hall conductivity to measure the OHE induced orbital accumulation in Cr thin films.
Our results conclusively demonstrate the ability to detect and quantify the OHE in a single non-magnetic layer
and provide direct evidence of a large current-induced orbital accumulation.
This opens a way to study the orbital Hall effect in metals
free of effects caused by an adjacent ferromagnet, which could bring new insights into the OHE, the SHE, and the connection between them.

%\section*{Acknowledgements}
\vspace{4pt}
We thank Shuyu Cheng, Katherine Robinson, Daniel Russell, and Fengyuan Yang for the growth of Pt samples.
We further thank Pavel Nov{\'a}k, J{\'a}n Rusz, and Markus Wei{\ss}enhofer for helpful comments.
I.L.\ and R.K.K.\ acknowledge support from the Center for Emergent Materials, an NSF MRSEC, under award number DMR-2011876.
This work was furthermore supported by the Swedish Research Council (VR), the Swedish National Infrastructure for Computing (SNIC) (Grant No.\ 2018-05973), and the K.\ and A.\ Wallenberg Foundation (Grant No.\ 2022.0079).

%\section*{Author contributions}

%All authors participated in data analysis and preparation of the manuscript.

%\newpage

\bibliography{OHE_Cr.bib}

\newpage

\end{document}